\documentclass[11pt,oneside]{article}
\usepackage{a4wide}
\usepackage{mathrsfs}
\usepackage{latexsym,bm}
\usepackage{graphicx}
\usepackage{indentfirst}
\usepackage{slashed}
\usepackage{amsmath}
\usepackage{amssymb}
\usepackage{color}
\usepackage[colorlinks=true,linkcolor=blue]{hyperref}
\usepackage{epsfig}
\usepackage[titletoc]{appendix}
\usepackage{multirow}%
\usepackage{rotating}
\usepackage{cite}
\usepackage{ulem} 
\usepackage[top=2.9cm,bottom=2.5cm,left=2.8cm,right=3cm]{geometry}
\usepackage{tabularx}

\newcommand{\email}[1]{\footnote{{\em } \texttt{#1}}}

\newcommand{\zc}{Z_c(3900)}
\newcommand{\zcs}{Z_{cs}(3985)}
\newcommand{\zca}{X(4020)}
\newcommand{\dbar}{\bar{D}}
\newcommand{\dvbar}{\bar{D}^{*}}
\newcommand{\dv}{D^{*}}
\newcommand{\dpn}{{D}^{0}}

\newcommand{\dvn}{D^{*0}}

\newcommand{\jpsi}{J/\psi}
\newcommand{\km}{K^{-}}
\newcommand{\hc}{h_c}
\newcommand{\ds}{D_{s}^{-}}
\newcommand{\dsv}{D_{s}^{*-}}

\newcommand{\vr}{\mathbf{r}}
\newcommand{\vk}{\mathbf{k}}

\begin{document}
\thispagestyle{empty}
\title{
\Large \bf A unified description of the hidden-charm tetraquark states $Z_{cs}(3985)$, $Z_c(3900)$, and $X(4020)$ }
\author{\small Zhi-Hui Guo$^{a,b}$\email{zhguo@seu.edu.cn}, \,  J.~A.~Oller$^{c}$\email{oller@um.es}  \\[0.3em]
{\small\it ${}^a$ School of Physics, Southeast University, Nanjing 211189, China } \\[0.1em]
{ \small\it ${}^b$  Department of Physics and Hebei Advanced Thin Films Laboratory, } \\
{\small\it Hebei Normal University,  Shijiazhuang 050024, China}\\[0.1em]
{\small {\it $^c$ Departamento de F\'{\i}sica. Universidad de Murcia. E-30071 Murcia. Spain}}
}
\date{}

%

\maketitle
\begin{abstract}
The newly observed hidden-charm tetraquark state $Z_{cs}(3985)$, together with $Z_c(3900)$ and $X(4020)$, are studied in the combined theoretical framework of the effective range expansion, compositeness relation and the decay width saturation. The elastic effective-range-expansion approach leads to sensible results for the scattering lengths, effective ranges and the compositeness coefficients, $i.e.$, the probabilities to find the two-charm-meson molecule components in the tetraquark states. The coupled-channel formalism by including the $J/\psi\pi$ and $D\bar{D}^*/\bar{D}D^*$ to fulfill the constraints of the compositeness relation and the decay width, confirms the elastic effective-range-expansion results for the $Z_c(3900)$, by using the experimental inputs for the ratios of the decay widths between $D\bar{D}^*/\bar{D}D^*$ and $J/\psi\pi$. With the results from the elastic effective-range-expansion study as input for the compositeness, we generalize the discussions to the $Z_{cs}(3985)$ by including the $J/\psi K^{-}$ and $D_s^{-}D^{*0}/D_s^{*-}D^{0}$, and predict the partial decay widths of the $J/\psi K^{-}$. Similar calculations are also carried out for the $X(4020)$ by including the $h_c\pi$ and $D^*\bar{D}^*$, and the partial decay widths of the $h_c\pi$ is predicted. Our results can  provide useful guidelines for future experimental measurements. 
\end{abstract}

\section{Introduction}

Recently, a prominent resonance state, named as $\zcs$, was discovered in the $\ds\dvn$/$\dsv\dpn$ invariant-mass distributions by the BESIII collaboration~\cite{Ablikim:2020hsk}. Most likely, it corresponds to the strange partner of the hidden-charm tetraquark states $Z_c(3900)$, which have been observed in the $\dbar\dv/D\dvbar$ and $\jpsi\pi$ event distributions~\cite{bibzc}. The observation of the $\zcs$ clearly constitutes a very important step toward the completion of the intriguing hidden-charm tetraquark spectra. Shortly after its discovery, there appear many interesting theoretical interpretations of the $\zcs$~\cite{bibzcs}, including  kinematical effects, a $\ds\dvn/\dsv\dpn$ molecular nature, compact $c\bar{c}s\bar{q}$ tetraquark meson, etc. 

Since the $\zc$  clearly manifests as a peak in the $\jpsi\pi$ invariant-mass distribution~\cite{bibzc}, it is natural to expect a similar enhancement in the $\jpsi \bar{K}$ channel for the $\zcs$. However, currently there are only experimental observations in the $\ds\dvn/\dsv\dpn$ channel for the $\zcs$~\cite{Ablikim:2020hsk}, and the $\jpsi \bar{K}$ channel is not mentioned in the experimental analysis yet, which is probably due to the limited statistics. This work aims at further clarifying the internal structure of the $\zcs$ and its possible decay width to the $\jpsi\km$ channel. We hope to provide useful results for future experimental measurements.

Weinberg's compositeness relation offers a valuable formalism to dissect the interior constituents of a hadron at the quantitative level~\cite{Weinberg:1962hj,Weinberg:1965zz}.\footnote{We give in Appendix~\ref{app.201219.1} an alternative general derivation of Weinberg's relation for nonrelativistic bound states.} Although it can give sensible and model-independent results for the bound-state case, the extension of the Weinberg's compositeness relation to the virtual and resonance poles is quite involved and still under severe debate~\cite{Baru:2003qq,Hanhart:2011jz,Hyodo:2011qc,Aceti:2012dd,Sekihara:2014kya,Guo:2015daa,Matuschek:2020gqe}. In Ref.~\cite{Guo:2015daa}, we have proposed to take phase transformations of the two-body scattering $S$  matrix to define the probabilistically meaningful compositeness coefficient, so that the latter is eligible to be interpreted as the probability of the two-hadron component in the resonance. This approach has been extensively applied to various resonances, including the light-flavor hadrons~\cite{Guo:2015daa} and many heavy exotic multiquark  candidates~\cite{Gao:2018jhk,Guo:2019kdc,Kang:2016jxw,Guo:2020pvt,Meissner:2015mza}.

A simple but effective coupled-channel method, based on the resonance compositeness formula derived in Ref~\cite{Guo:2015daa} and the decay width saturation, is introduced in Ref.~\cite{Meissner:2015mza} to study the hidden-charm pentaquark baryons. This approach  has been shown to be 
quite useful in providing important information on the possible exotic hadrons with very few inputs~\cite{Guo:2019kdc,Guo:2020pvt}, which are the mass and width of the resonance and the total compositeness coefficient ($X$). The latter is an external input, which should be provided beforehand if not enough knowledge on the partial-decay widths is available. For instance, a coupled-channel calculation of the partial-wave amplitudes involved  was used  to calculate $X$ in Ref.~\cite{Guo:2020pvt} for the $X(6900)$. Due to the close proximity of the $\zcs$ to the open charm $\ds\dvn/\dsv\dpn$ thresholds, it is natural to assume that the compositeness of the hadronic degrees of freedom is dominated by the $\ds\dvn/\dsv\dpn$ for the $\zcs$. The single-channel compositeness coefficient can be then calculated by the elastic effective-range-expansion (ERE) formulation~\cite{Guo:2016wpy,Kang:2016ezb,Oller:2017alp}. The dominance of the compositeness coefficient by the nearby threshold will be explicitly verified in the $\zc$ case by using the experimental ratios of its decay widths to $\jpsi\pi$ and $\bar{D}\dv/D\dvbar$. This also gives us confidence to further rely on the compositeness coefficient from the elastic ERE approach to estimate the partial decay widths for other related resonances such as the $\zca$ and $\zcs$. To be more specific, we include the $\jpsi K^{-}$ and $\ds\dvn/\dsv\dpn$ coupled channels to study the $\zcs$, and consider the $\hc\pi$ and $\dv\dvbar$ to describe the $\zca$.

The contents of this work are organized as follows. First, we elaborate in Sec.~\ref{sec.201221.1} on the elastic ERE study of the two-charm-meson scattering, where the scattering lengths, effective ranges and compositeness coefficients are provided. In the next Sec.~\ref{sec.201221.2}, a complementary and more robust coupled-channel framework is used to address the partial decay widths and partial compositeness coefficients.  Finally, we give a short summary and conclusions in Sec.~\ref{sec.201221.3}. In addition, some extra related material has been included in the Appendix~\ref{app.201219.1} on the compositeness relation and in applying the exactly solvable model of a source in Quantum Field Theory to relate the radiated real particles with the cloud of  virtual ones surrounding a resonant source. 

\section{Elastic two-charm-meson scattering in the effective range expansion} 
\label{sec.201221.1}

A featured property shared by many of the recently observed heavy-flavor exotic states is that most of them lie near some specific thresholds of the underlying two hadronic states. The effective range expansion, relying on the three-momentum expansion at the two-body threshold, naturally provides a useful framework to investigate the physics around the threshold energy region. Up to next-to-leading order, there are two parameters in the elastic ERE expansion, namely the scattering length $a$ and effective range $r$. Combining the ERE expansion and  unitarity, one can uniquely determine $a$ and $r$ in terms of the mass and the width of the resonance~\cite{Guo:2016wpy,Kang:2016ezb,Hyodo:2013iga}.

For the two-hadron $S$-wave scattering, the single-channel unitarized ERE formula up to next-to-leading order has the form  
\begin{eqnarray}\label{eq.ti}
T(E)=\frac{1}{-\frac{1}{a}+\frac{1}{2}r\,k^2-i\,k}\,,
\end{eqnarray}
with $k$ being the center-of-mass (CM) three-momentum. The nonrelativistic relation between the energy $E$ and three-momentum $k$ reads  
\begin{eqnarray}
k= \sqrt{2\mu_m(E-m_{\rm th})}\,,
\end{eqnarray}
with $\mu_m=\frac{m_1m_2}{m_1+m_2}$, $m_{\rm th}=m_1+m_2$ and $m_1$ and $m_2$ the masses of the scattering states. The elastic ERE amplitude of Eq.~\eqref{eq.ti} obeys the unitarity condition,
\begin{equation}
 {\rm Im}\,T(E)^{-1} = - k  \,,\quad (E > m_{\rm th})\,.
\end{equation}
The virtual and resonance poles are located in the second Riemann sheet (RS), in which the ERE  amplitude is given by 
\begin{eqnarray}\label{eq.tii}
 T^{\rm II}(E)=\frac{1}{-\frac{1}{a}+\frac{1}{2}r\,k^2+i\,k}\,.
\end{eqnarray}
It should be noted that the imaginary part of the $k$ should be positive in the  formulations~\eqref{eq.ti} and \eqref{eq.tii}. The resonance pole corresponds to the zeros of the denominator of $T^{\rm II}(E)$.
After straightforward algebraic manipulations, one can derive the following expressions for $a$ and $r$ 
in order to reproduce a resonance pole at $E_R=M_{R}-i\Gamma_R/2$ \cite{Kang:2016ezb}, 
\begin{eqnarray}\label{eq.ar}
 a=-\frac{2k_{\rm i}}{k_{\rm r}^2+k_{\rm i}^2}\,, \qquad r= -\frac{1}{k_{\rm i}}\,,
\end{eqnarray}
where the real ($k_{\rm r}$) and imaginary ($k_{\rm i}$) parts of the pole position in the $k$ variable are defined as 
\begin{eqnarray}\label{eq.kr}
 k_R=\sqrt{2\mu_m(E_R-m_{\rm th})}\equiv k_{\rm r} + i k_{\rm i}\,,\quad (k_{\rm i}>0)\,. 
\end{eqnarray}

The  Laurent expansion of $T^{\rm II}(E)$ around the pole position is 
\begin{eqnarray}\label{eq.tiilr}
T^{\rm II}(E) = \frac{\gamma_k^2}{k-k_R} + \cdots \,,
\end{eqnarray}
where the ellipsis corresponds to the regular terms in the $k-k_R$ expansion. Combining Eqs.~\eqref{eq.tii}-\eqref{eq.tiilr}, it is straightforward to derive the relation between the residue $\gamma_k^2$ and the pole position 
\begin{eqnarray}\label{eq.gammak}
\gamma_k^2= -\frac{k_i}{k_r}\,.
\end{eqnarray}
The compositeness coefficient $X$, i.e. the probability to find the two-body component in the resonance, comes as an important byproduct from the unitarized ERE amplitude, which turns out to be~\cite{Kang:2016ezb}
\begin{eqnarray}\label{eq.x}
 X=\gamma_k^2=-\frac{k_i}{k_r}~,
\end{eqnarray}
which is $\leq 1$ for $M_R\geq m_{\rm th}$ \cite{Kang:2016ezb}. 
We show in the Appendix~\ref{app.201219.1} how the previous formula is connected with the direct calculation of the weight of a channel in a nonrelativistic bound state by calculating the expectation value of the number operator of free particles in the state, following Ref.~\cite{Oller:2017alp}. Its extension to resonances is illustrated with a exactly solvable toy-model in Quantum Field Theory by including a source that mimics a decaying resonance.

In terms of $a$ and $r$, Eq.~\eqref{eq.x} can be cast as 
\begin{eqnarray}\label{eq.xb}
 X=\left(\frac{2r}{a}-1\right)^{-\frac{1}{2}}\,.
\end{eqnarray}
According to Eqs.~\eqref{eq.ar} and \eqref{eq.x}, the scattering length, effective range and the compositeness coefficient can be obtained, once the mass and width of the resonance are provided. The constraint $0\le X \le 1$ requires that i) $r/a\geq 1/2$, as it trivially follows from Eq.~\eqref{eq.ar}, and ii) $r\le a$ (or $|r|\ge|a|$), which implies that $M_R\geq m_{\rm th}$ as derived in detailed in Ref.~\cite{Kang:2016ezb}. In the general case when there are other types of poles, such as virtual and bound states, $a$ and $r$ could take different relative sizes and signs~\cite{Ikeda:2011dx}. Our Equation~\eqref{eq.xb} is not intended to cope in such cases that do not enter in our specific study here. However, in the literature one can find different  approaches  for the virtual and bound states in Refs.~\cite{Weinberg:1965zz,Baru:2003qq,Matuschek:2020gqe,Hyodo:2013iga}.

We now apply the elastic ERE formalism to the $S$-wave $D\dvbar/\bar{D}\dv$, $\dv\dvbar$ and $\dsv\dpn/\ds\dvn$ scattering processes, which are expected to generate resonance poles for the $\zc$, $\zca$ and $\zcs$, in this order.
The analysis for the case of the $\zcs$ requires some clarifying remarks.
Though the two channels $\dsv\dpn$ and $\ds\dvn$ are distinguishable, they both scatter in $S$ wave and their threshold is only around 2 MeV apart, a separation that is much smaller than the width of the $\zcs$. Hence, for simplicity in the calculations, we take both thresholds to be coincident, either with the value of one or the other threshold.
Performing an elastic ERE study is possible because the ERE expansion of a $2\times 2$ partial-wave amplitude with equal thresholds can be diagonalized by a real orthogonal matrix. We then apply our elastic ERE study to  the resonance eigenchannel.

The current experimental analyses cannot distinguish between the $\dsv\dpn$ and $\ds\dvn$ states in the event distributions.
In order to verify the differences of the two channels, we separately calculate the $a$, $r$ and $X$ by taking the threshold of either $\dsv\dpn$ or $\ds\dvn$.
The results from the two channels turn out to be compatible within uncertainties, as shown in last two rows of Table~\ref{tab.ar}.
In the same table we also give for completeness  $a$, $r$ and $X$ for the $\dbar\dv/D\dvbar$ and $\dv\dvbar$ scattering corresponding to the resonances  $\zc$, $\zca$, respectively~\cite{Gao:2018jhk}.
According to the $X$ values in Table~\ref{tab.ar}, 
the two-charm-meson molecular components in the $\zc$, $\zca$ and $\zcs$ resonances are sizable, at the level around 40\%.
It is also interesting to point out that the nearby two-charm-meson molecular constituents for $\zc$, $\zca$ and $\zcs$ are rather similar.

The resulting magnitudes of the scattering lengths for all the three cases are around 1~fm. The magnitudes of the effective ranges $r$ are found to be around $3 \sim 4$~fm in Table~\ref{tab.ar},  larger than the typical QCD hadronic scale at 1~fm. According to the findings in Refs.~\cite{Guo:2016wpy,Kang:2016ezb}, in the special situation when there is a Castillejo-Dalitz-Dyson (CDD)  on top of the threshold $m_{\rm th}$, one would have $r \propto 1/(M_{\rm CDD} - m_{\rm th})^2$, i.e. a diverging effective range clearly indicates an underlying CDD pole in the system. The large values of  $|r|$ in Table~\ref{tab.ar} could be indicative that indeed this is the case. 
This indication seems to be also consistent with the moderate values for the total compositeness $X$.  The closely similar values for all the three resonances of $X$ and $r$, and even of $a$, seem supporting the conjecture that the $\zc$, $\zca$ and $\zcs$ form the hidden-charm multiplet, with an important component of  the 
charm meson pairs $D_{(s)}^{(*)}\bar{D}_{(s)}^{(*)}$ together with a bare component, that seems equally important. 
A future experimental measurement of the $D_{s}^{(*)}\bar{D}_{s}^{(*)}$ line shapes will be helpful to assure the conjecture.

\begin{table}[htbp]
\centering
\begin{scriptsize}
\begin{tabular}{ c c c c c c c}
\hline\hline
 Tetraquark & Mass   & Width & Threshold  & $a$     & $r$   &  $X$  \\
  Resonance         & (MeV)  & (MeV) &  (MeV)    & (fm)    &  (fm) &   
\\ \hline
$\zc$  & $3888.4\pm 2.5$ & $28.3\pm 2.5$ & $\dbar\dv$~(3875.5)  & $-0.84 \pm 0.13$ & $-2.52 \pm 0.25$  & $0.45\pm 0.06$
\\ \hline
$\zca$ & $4024.1\pm 1.9$ & $13\pm 5$   & $\dvbar\dv$~(4017.1)   & $-1.04 \pm 0.30$ & $-3.90 \pm 1.35$  & $0.39\pm 0.14$ 
\\ \hline
$\zcs$ & $3982.5\pm 3.3$ & $12.8\pm 6.1$ & $\ds\dvn$~(3975.2)  & $-1.00 \pm 0.47$ & $-4.04 \pm 1.82$ & $0.38\pm 0.18$ \\
       &   &                             & $\dsv\dpn$~(3977.0) & $-1.28 \pm 0.60$ & $-3.65 \pm 1.60$ & $0.46\pm 0.19$ 
\\\hline\hline
\end{tabular}
\end{scriptsize}
\caption{\label{tab.ar} Values of $a$, $r$ and the total compositeness $X$ for the resonances $\zc$, $\zcs$ and $\zcs$ from the elastic ERE study  taking as input the mass and width of every resonance, which are given in the second and third columns, respectively. For the $\zcs$ the analysis is done twice by taking either the threshold of $\ds\dvn$ or $\dsv\dpn$.} 
\end{table}

\section{Coupled-channel study of the hidden-charm tetraquark candidates}
\label{sec.201221.2}

A more realistic study of the near-threshold hidden-charm tetraquark state requires the coupled-channel calculation. Since the $\zc$ and $\zca$ are experimentally observed in the $\jpsi\pi$, $D\dvbar$ and $\hc\pi$, $\dv\dvbar$ event distributions, respectively, it is natural to include the $\jpsi\pi$ and $D\dvbar$ coupled channels to study the $\zc$, and the $\hc\pi$ and $\dv\dvbar$ to address the $\zca$. For the $\zcs$, the $\jpsi\km$ and $\ds\dvn/\dsv\dpn$ channels will be considered.

The coupled-channel formalism adopted here relies on the decay width saturation and the compositeness relation, which has been successfully used to describe the hidden-charm pentaquark candidates~$P_c(4312)$, $P_c(4440)$ and $P_c(4457)$~\cite{Meissner:2015mza,Guo:2019kdc}, and the fully charmed tetraquark state $X(6900)$~\cite{Guo:2020pvt}. After a proper phase transformation of the $S$ matrix, we have showed in Ref.~\cite{Guo:2015daa} that the partial  compositeness coefficient $X_j$ contributed by the $j$th channel to a resonance can be written as 
\begin{eqnarray}
\label{201218.4}
X_j = |g_j|^2 \, \bigg|\frac{\partial G_{j}^{\rm II}(s_R)}{\partial s}\bigg| \,,
\end{eqnarray}
where $s=E^2$ is the usual Mandelstam variable, $s_R=E_R^2$ with $\Re s_R>m_{\rm th}^2$, 
and the couplings $g_j^2$ are the residue of the partial-wave amplitude at the
pole position,
$\lim_{s\to s_R}(s-s_R)T^{\rm{II}}(s)_{jj}=-g_j^2$, cf. Eq.~\eqref{eq.gammak}. 
The function $G_j^{\rm II}(s)$ corresponds to the unitarity loop function $G(s)$ evaluated for the $j$th channel in the second RS. Due to the presence of the lighter threshold that is distant from the resonance pole, we will adopt the relativistic kinematical relations in the coupled-channel case. An explicit relativistic expression for the $G(s)$ function from the dimensional regularization is given by 
\begin{eqnarray}\label{eq.gfunc}
G(s)  &=& -\frac{1}{16\pi^2}\left[ a_{\rm SC}(\mu^2) + \log\frac{m_2^2}{\mu^2}-x_+\log\frac{x_+-1}{x_+}
-x_-\log\frac{x_--1}{x_-} \right]\,, \nonumber\\
 x_\pm &=&\frac{s+m_1^2-m_2^2}{2s}\pm \frac{q(s)}{\sqrt{s}}\,,
\end{eqnarray}
where $q(s)$ is the relativistic three-momentum
\begin{eqnarray}
  q(s) =\frac{\sqrt{[s-(m_{1}+m_{2})^2][s-(m_{1}-m_{2})^2]}}{2\sqrt{s}}\,,
\end{eqnarray}
and $m_1$ and $m_2$ are the masses of the two particles involved in the channel of interest. 
After taking the derivative of the $G(s)$ function, the unknown constant $a_{\rm SC}(\mu)$ disappears  and hence does not enter the compositeness coefficient $X$ in Eq.~\eqref{201218.4}. The total compositeness relation in the two-channel case has the form 
\begin{eqnarray}\label{eq.fx}
  X= X_1 + X_2 \equiv |g_1|^2 \, \bigg|\frac{\partial G_{1}^{\rm II}(s_R)}{\partial s}\bigg| +  |g_2|^2 \, \bigg|\frac{\partial G_{2}^{\rm II}(s_R)}{\partial s}\bigg|\,,
\end{eqnarray}
where the lighter channel is labeled as $1$ and the heavier two-charm-meson channel is labeled as $2$. 
The coupling $g_2^2$ is related to the residue in Eq.~\eqref{eq.gammak} via 
\begin{eqnarray}\label{eq.resigg}
|g^2_2|=\frac{16\pi|E_R^2 k_R|}{\mu_m}~. 
\end{eqnarray}

For the lighter channel, whose threshold is distant from the resonance mass, its partial decay width is calculated by the standard formula 
\begin{eqnarray}\label{eq.gamma1}
 \Gamma_1 =  |g_1|^2\frac{q_1(M_R^2)}{8\pi M_R^2}\,,
\end{eqnarray}
with $q_1(M_R^2)$ the CM three-momentum of the lighter channel. Due to the closeness of the resonance mass to the two-charm-meson threshold, the standard formula for the decay width becomes inadequate.
Instead, the Lorentzian energy distribution for the resonance mass is used to estimate the partial width $\Gamma_2$ for the two-charm-meson channel 
\begin{eqnarray}\label{eq.gamma2}
\Gamma_2= |g_2|^2 \int_{m_{\rm th}}^{M_R+n\,\Gamma_R} dE \,\frac{q_2(E^2)}{16\pi^2 \,E^2} \frac{\Gamma_R}{(M_R-E)^2+\frac{\Gamma_R^2}{4}}  \,,
\end{eqnarray}
which naturally reduces to the standard decay formulation in Eq.~\eqref{eq.gamma1} for the narrow-width resonance, i.e. when taking $\Gamma_R\to 0$. In principle the upper integration limit in Eq.~\eqref{eq.gamma2} should be taken to infinity. In Ref.~\cite{Kang:2016ezb}, the value of $n=8$ is estimated for the $Z_b(10610)/Z_b(101650)$ by reproducing the experimental widths using Eq.~\eqref{eq.gamma2}, with the coupling from the elastic ERE study. Following the same procedure for the $\zc$, the value of $n$ is found to be quite large around 60. Nevertheless, it is pointed out that the resulting width from Eq.~\eqref{eq.gamma2} with $n=10$ is already quite close to the experimental value, with the deviation less than 7 percent. In practice, we will take $n=10$ throughout by taking into account the relatively narrow widths of the $\zc$, $\zca$ and $\zcs$. Increasing the values of $n$  by 1 or 2 orders of magnitude  the resulting integral in Eq.~\eqref{eq.gamma2} changes less than 10 percent, so that the results and conclusions change  little by extending the upper integral limit in Eq.~\eqref{eq.gamma2}. In particular, they are well inside the estimated error uncertainty in the results given.

The requirement of the resonance width saturation gives 
\begin{eqnarray}\label{eq.fw}
 \Gamma_R = \Gamma_1 + \Gamma_2  = |g_1|^2\frac{q_1(M_R^2)}{8\pi M_R^2}+ |g_2|^2 \int_{m_{\rm th}}^{M_R+n\,\Gamma_R} dE \,\frac{q_2(E^2)}{16\pi^2 \,E^2} \frac{\Gamma_R}{(M_R-E)^2+\frac{\Gamma_R^2}{4}}\,. 
\end{eqnarray}

Combining Eqs.~\eqref{eq.fx} and \eqref{eq.fw} allows one to solve $|g_1|$ and $|g_2|$, with which one can obtain the valuable information on the partial decay widths and individual compositeness coefficients $X_i$ for every channel.
However, in order to solve Eqs.~\eqref{eq.fx} and \eqref{eq.fw} we need first to provide the total compositeness coefficient $X$, whose value is difficult to estimate beforehand.
Different ways to solve the equations are proposed in the previous studies.
For example, for the pentaquark candidates $P_c$, some arbitrary testing values between 0 and 1 are taken for the $X$~\cite{Guo:2019kdc}.
For the fully charmed tetraquark candidate $X(6900)$, a sophisticated dynamical model respecting  unitarity is constructed to fit the experimental line shapes first, so that the couplings $|g_j|$ are then extracted from the dynamical coupled-channel amplitudes.
It is worth pointing out that the neat recipe of the coupled-channel method of Eqs.~\eqref{eq.fx} and \eqref{eq.fw} agrees perfectly with the sophisticated coupled-channel amplitudes~\cite{Guo:2020pvt}. 

Regarding the $\zc$, the ratio of the decay widths between the $D\dvbar$ and $\jpsi\pi$ is measured to be $6.2\pm 2.9$ by the BESIII experiment~\cite{bibzc}.
Indeed the ratio of the decay widths also provides another key input to solve the coupled-channel systems~\eqref{eq.fx} and~\eqref{eq.fw}, so that now it is not necessary to fix the total compositeness $X$ $a$ $priori$, and it can be predicted. With the experimental input, 
\begin{align}
\label{201218.1}
\Gamma_{D\dvbar}/\Gamma_{\jpsi\pi}=6.2\pm 2.9~,
\end{align}
it is easy to obtain from Eq.~\eqref{eq.fw} that 
\begin{eqnarray}
\label{201218.2}
|g_1|= 1.46_{-0.23}^{+0.43}\,, \qquad |g_2|= 7.89_{-0.44}^{+0.18}\,.
\end{eqnarray}
Hence, by implementing these values into Eq.~\eqref{eq.fx} one has that  
\begin{eqnarray}\label{eq.zcx}
X_1=0.002\pm0.001\,, \qquad X_2= 0.436_{-0.047}^{+0.021}\,, \qquad X=X_1+X_2= 0.438_{-0.047}^{+0.021}\,. 
\end{eqnarray}
Now it is interesting to compare the coupled-channel predictions in Eq.~\eqref{eq.zcx} with the elastic ERE result in Table~\ref{tab.ar}. The total compositeness coefficient $X$, which is overwhelmingly dominated by the near-threshold $D\dvbar/\dbar\dv$ channel, is almost identical as the value from the elastic ERE study. That is, the ERE study reproduces the coupled-channel analysis because the role played by the lightest channel is rather marginal, as reflected by the small values of $|g_1|$ and $X_1$ compared to those of $|g_2|$ and $X_2$ in Eqs.~\eqref{201218.2} and \eqref{eq.zcx}, respectively.

An alternative way to estimate the compositeness $X_2$ is to rewrite the partial decay width~\eqref{eq.gamma2} as 
\begin{eqnarray}\label{eq.gamma2x}
\Gamma_2= \frac{X_2 |k_R| \,\, |E_R|^2 }{\pi \mu_m} \int_{m_{\rm th}}^{M_R+n\,\Gamma_R} dE \,\frac{q_2(E^2)}{E^2} \frac{\Gamma_R}{(M_R-E)^2+\frac{\Gamma_R^2}{4}}  \,,
\end{eqnarray}
which is obtained by substituting Eqs.~\eqref{eq.x} and \eqref{eq.resigg} into Eq.~\eqref{eq.gamma2}. By denoting the branching ratio of the decay width of the resonance $R$ into the two-charm-meson channel as $b_2$, i.e. $\Gamma_2=b_2\,\Gamma_R$, the above equation then leads to 
\begin{eqnarray}\label{eq.x2}
X_2= \frac{\pi \mu_m b_2  }{|k_R| \,\, |E_R|^2 \int_{m_{\rm th}}^{M_R+n\,\Gamma_R} dE \,\frac{q_2(E^2)}{E^2} \frac{1}{(M_R-E)^2+\Gamma_R^2/4}}   \,.
\end{eqnarray}
By assuming the width saturation of the $\zc$ by the $\jpsi\pi$ and $D\dvbar$ channels and taking into account the experimental ratios of their decay widths, our prediction from Eq.~\eqref{eq.x2} is 
\begin{equation}\label{eq.x2v}
X_2=0.41_{-0.04}^{+0.02}\,,
\end{equation}
in which uncertainties are purely from the experimental ratios. The two different predictions in Eqs.~\eqref{eq.zcx} and \eqref{eq.x2v}
are clearly compatible with each other within statistical uncertainties (as they should be).

Based on the similarities of the scattering lengths, effective ranges and compositeness coefficients for the three tetraquark candidates $\zc$, $\zca$ and $\zcs$, as shown in Table~\ref{tab.ar}, it seems justified to estimate the total compositeness coefficients in the coupled-channel systems by using the ones from the elastic ERE for the $\zca$ and $\zcs$. The consistency with this assumption requires that the coupling and partial compositeness for the lighter channel in every resonance are much smaller than those for the heavier one, as we have shown for the $\zc$, and, indeed, this is the case. 
The coupled-channel solutions for the $\zca$ and $\zcs$ are summarized in Table~\ref{tab.xzcas}. For both resonances, apart from taking the $X$ from the elastic ERE study ($X_{\rm ERE}$), we have also tried using other different total compositeness values.  Generally speaking, the solutions for the $\zca$ and $\zcs$ share the same features. The smaller total compositeness coefficients lead to larger partial widths for the light channels and smaller widths for the two-charm-meson channels \cite{Meissner:2015mza}. In all the cases, for $X_{\rm ERE}$ the compositeness  and width are dominated by the near-threshold two-charm-meson components and the approach is consistent, while showing a remarkable similar pattern in the properties of the resonances $\zc$, $\zca$ and $\zcs$.

Our prediction for the partial decay width of the $\zcs$ into the $\jpsi K^-$ channel could provide a useful guide for the future experimental analysis. Similarly, up to now the decay ratio to the $\dv\dvbar$ and $\hc\pi$ from the $\zca$ is not available from experiment, being that the former is predicted to be much larger than the latter (with a ratio of widths of around 8). Future measurement of this quantity could be an important criterion to judge the theoretical formalism proposed in this work and conclude whether the  $\zc$, $\zca$ and $\zcs$ form the hidden-charm multiplet.

\begin{table}[htbp]
\centering
\begin{scriptsize}
\begin{tabular}{ c c c c c c c }
\hline\hline
Resonance & $|g_1|$ & $|g_2|$   & $\Gamma_1$ & $\Gamma_2$  & $X_1\times 10^{3}$   & $X_2$    \\
       &  (GeV)  & (GeV)     & (MeV)     &  (MeV)      &         &        \\
\hline
$\zca$    &   &      &      &        &         &        \\
$X_\text{ERE}=0.39\pm 0.14$ & $1.1 \pm 0.2 $ & $6.5\pm 1.3$ & $1.4\pm 0.5$  & $11.6\pm 4.5$ & $1\pm 1$ &  $0.39\pm 0.14$\\
\hline
$\zcs$ &   &    &  &   &   &   \\
 \hline
Threshold$({\ds\dvn})$ &  &      &     &       &         &     \\
$X_{\text{ERE}}=0.38\pm 0.18$ & $0.8\pm 0.2$ & $6.4\pm 1.7$ & $1.2\pm 0.6$  & $11.6\pm 5.3$ & $0.8\pm 0.4$ &  $0.38\pm 0.18$\\
 \hline
Threshold$({\dsv\dpn})$ &  &      &     &       &         &      \\
$X_{\text{ERE}}=0.46\pm 0.19$  & $0.9\pm 0.2$ & $6.8\pm 1.7$ & $1.2\pm 0.6$  & $11.6\pm 5.6$ & $0.8\pm 0.4$ &  $0.46\pm 0.19$\\
\hline\hline
\end{tabular}
\end{scriptsize}
\caption{\label{tab.xzcas} The coupled-channel solutions for the $\zca$ and $\zcs$.
  For the $\zca$, the subscripts 1 and 2 stand for the channels $\hc\pi$ and $\dv\dvbar$, respectively, and for the 
$\zcs$ they refer to the $\jpsi K^-$ and $\ds\dvn/\dsv\dpn$ channels, in this order.  
See the text for further details. } 
\end{table}

\section{Summary and conclusions}
\label{sec.201221.3}

In this work we have studied the three hidden-charm tetraquark candidates $\zc$, $\zca$ and the newly observed $\zcs$. The $D\dvbar/\dbar\dv$, $\dv\dvbar$ and $\ds\dvn/\dsv\dpn$ channels are incorporated to investigate the $\zc$, $\zca$ and $\zcs$, respectively, within the elastic effective-range-expansion scheme. The scattering lengths $a$, effective ranges $r$,  and compositeness coefficients $X$, $i.e.$, the probabilities to find the two-charm-meson constituents in the resonances, are calculated. The rather large magnitudes of $r$ found, between 3 and 4 fm, hint toward the relevance of  possible underlying bare CDD poles, which seems consistent with the moderate compositeness $X$ obtained.
Therefore, the elastic effective-range-expansion study reveals that both the two-charm-meson molecular components and other degrees of freedom, such as the compact four-quark cores or heavier hadronic components, would play relevant roles in the physical $\zc$, $\zca$ and $\zcs$ resonances. 

The elastic effective-range-expansion description is then improved by the coupled-channel formalism, which is based on the simultaneous saturation of the compositeness and decay width. 
By using the ratio of the decay widths between the $D\dvbar/\dbar\dv$ and $\jpsi\pi$ channels from the BESIII experiment, we exactly solve the coupled-channel system for the $\zc$. The compositeness value $X$ turns out to be almost the same as the one from the elastic effective range expansion. The coincidence of the elastic study with the coupled-channel one reflects the marginal role played by the lighter channel, as evidenced by its small coupling and the related partial compositeness. Guided by this finding in the $\zc$ study,  we then predict the partial widths to the $\hc\pi$ and $\dv\dvbar$ channels for the $\zca$ and similarly for the $\zcs$ into $\jpsi K^-$ and $\ds\dvn/\dsv\dpn$. As in the case of the $\zc$, the decay widths to the lighter channels are predicted to be much smaller than those to the heavier ones. We also point out that the statistical uncertainty in our results could be  improved by more precise measurements of the total widths of the resonances mentioned. We think that our results can provide helpful guidelines for future measurements in relevant experiments.

\section*{Acknowledgements}
This work is partially funded by the Natural Science Foundation of China under Grant Nos.~11975090 and ~11575052, the Natural Science Foundation of Hebei Province under Contract No.~A2015205205, the Fundamental Research Funds for the Central Universities, the MINECO (Spain) and EU grant FPA2016-77313-P and the MICINN (Spain) grant PID2019-106080GB-C22.

\appendix

\section{Calculation of the compositeness for a non-relativistic bound state}
\setcounter{equation}{0}
\label{app.201219.1}
\def\theequation{\Alph{section}.\arabic{equation}}

In Ref.~\cite{Oller:2017alp}, the partial compositeness $X_A$ for a particle species $A$ of a nonrelativistic bound state was shown to be given by the expectation value of the corresponding number operator of free particles of type $A$ divided by its maximum number $n_A$. For example, in the case of the deuteron the compositeness of nucleons is the average number of nucleons divided by 2. For simplicity in the writing we suppress the subscript $A$ and just refer to $X$.

We reproduce here the formula Eq.~(39) of Ref.~\cite{Oller:2017alp} for calculating $X$ in terms of the coupling squared $g^2(k^2)$ of the bound state to this channel as a function of the three-momentum $k^2$, 
\begin{align}
\label{201218.1a}
X&=\frac{2\mu_m^2}{\pi^2}
\int_{0}^{+\infty}dk \,k^2 \frac{g^2(k^2)}{(k^2-\kappa^2)^2}~,\\
\kappa^2&=2\mu_m E_B~,\nonumber
\end{align}
with $E_B$ the (negative) binding energy. We next perform a Taylor expansion of the coupling squared around $\kappa^2$ and calculate the resulting integrations in dimensional regularization, in which, the powerlike divergences vanish. The output of this calculation yields the interesting formula
\begin{align}
\label{201218.2a}
X=-g^2(\kappa)^2\frac{\partial G(E_B)}{\partial E_B}-\frac{\partial g^2(\kappa^2)}{\partial \kappa^2}\frac{\mu^2|\kappa|}{\pi}~,
\end{align}
with the nonrelativistic unitarity loop function $G(E)$ given by
\begin{align}
\label{201218.3}
G(E)&=\int_0^\infty \frac{dk\,k^2}{2\pi^2}\frac{1}{E-k^2/2\mu_m}~. 
\end{align}
If $\Lambda^{-1}$ is the typical range of the finite range interactions, due to the onset of the left-hand cut because of crossed particle exchanges, the second term on the right-hand side of Eq.~\eqref{201218.2a} is ${\cal O}(\kappa^2/\Lambda^2)$ compared to the first one.\footnote{This approximation is typically much better for resonances in the heavy-quark hadrons because pion exchanges are perturbative \cite{valderrama,vanKolck} and/or because of OZI rule suppression of light meson exchanges like in the $X(6900)$ \cite{Guo:2020pvt}.} Thus, for close enough bound states to the threshold we can approximate $X$ by just the first term,
\begin{align}
\label{201218.2b}
X=-g^2(\kappa)^2\frac{\partial G(E_B)}{\partial E_B}+{\cal O}\left(\frac{\kappa^2}{\Lambda^2}\right)=i\gamma_k^2+{\cal O}\left(\frac{\kappa^2}{\Lambda^2}\right)~,
\end{align}
which constitutes an alternative general derivation of the well-known Weinberg's formula for $X$ \cite{Weinberg:1965zz}, indicated in this paper by $1-Z$, with $Z$ the elementariness.  
When this formula is continued to a resonance pole it implies Eq.~\eqref{eq.x}, 
once  its modulus is taken \cite{Guo:2015daa,Kang:2016ezb}, as e.g. explicitly indicated in Eq.~\ref{201218.4}. 

We would also like  to develop a simple toy-model example to illustrate the close relationship between the virtual particles in the cloud of a source at early times and the real ones that are radiated at later times. As a result, we think that this toy model shows that counting the later ones gives then information about the virtual cloud particles in the composition of a resonance. The example is based on an exactly solvable model in Quantum Field Theory \cite{thirring.181101.1} consisting of a source coupled to a scalar field $\phi(x)$. The Lagrangian density is
\begin{align}
\label{201218.5}
{\cal L}&=\frac{1}{2}\partial_\mu\phi\partial^\mu\phi-\frac{m^2}{2}\phi^2+g\rho(x)\phi~,
\end{align}
where $\rho(x)$ is a source (external field). To mimic a decaying resonance of mass $M_R$ and width $\Gamma$ we consider a source that is produced at $t=0$ and that later experiences damping oscillations. Namely,
\begin{align}
\label{201218.6}
\rho(x)&=\left\{
\begin{array}{lr}
  g\rho(\vr) \cos(M_R t) e^{-\frac{\Gamma t}{2}}& t>0 \\
  0         & t<0
\end{array}
\right.
\end{align}
with $g$ a constant coupling and the source can be taken without loss of generality to be normalized to one ($\int d^3r \rho(\vr)=1$). 
The Fourier transform of this source in space and time is
\begin{align}
\label{201218.7}
\rho(k)&=\frac{ig\rho(\vk)(k^0+i\Gamma/2)}{(k^0-M_R+i\Gamma/2)(k^0+M_R+i\Gamma/2)}~,\\
\rho(\vk)&=\int d^3r e^{-i\vk\vr}\rho(\vr)~.\nonumber
\end{align}
Making use of the advanced Green function $\Delta^{\rm adv}(x)$ the field $\phi(x)$ can be written as
\begin{align}
\label{201218.8}
\phi(x)&=\phi^{\rm out}(x)+\int d^4x'\Delta^{\rm adv}(x,x')g\rho(x')~,
\end{align}
where $\phi^{\rm out}(x)$ is the field at the far future ($t\to +\infty$) expressed in terms of the asymptotic creation and annihilation operators $B(\vk)^{(\dagger)}$ of the real free particles.
Performing explicitly the integration over $x'$ in Eq.~\eqref{201218.8} one finds that
\begin{align}
\label{201218.9}
\phi(x)&=\phi^{\rm out}(x)+g\int\frac{d^4k}{(2\pi)^42w(k)}e^{-ikx}\rho(k)
\left(\frac{1}{k^0+w(k)-i\alpha}-\frac{1}{k^0-w(k)-i\alpha}\right)~,
\end{align}
with $w(k)=\sqrt{\vk^2+m^2}$ and $\alpha\to 0^+$. From this formula one can readily obtain the relationship between
the annihilation operators $a(\vk,t=0)$ of the virtual particles (in terms of which $\phi(\vr,t=0)$
is expressed as if it were a free field only at $t=0$) and $B(\vk)$.
Using $\rho(k)$ from Eq.~\eqref{201218.7} it reads
\begin{align}
\label{201218.9a}
a(\vk,t=0)&=B(\vk)+g\rho(\vk)\frac{w(k)-i\Gamma/2}{(w(k)+M_R-i\Gamma/2)(w(k)-M_R-i\Gamma/2)}~.
\end{align}
Then, the expression for the total number of virtual particles $\bar{n}$
is\footnote{This calculation follows by noticing that $\displaystyle{a(\vk,t=0)|0\rangle_{\rm out}=
  g\rho(\vk)\frac{w(k)-i\Gamma/2}{(w(k)+M_R-i\Gamma/2)(w(k)-M_R-i\Gamma/2)}}|0\rangle_{\rm out}$, where $|0\rangle_{\rm out}$ is the outgoing ground state annihilated by the $B(\vk)$.}
\begin{align}
\label{201218.9b}
\bar{n}&=\int\frac{d^3k}{(2\pi)^32w(k)}\frac{|g\rho(\vk)|^2(w(k)^2+\Gamma^2/4)}{[(w(k)+M_R)^2+\Gamma^2/4][(w(k)-M_R)^2+\Gamma^2/4]}~.
\end{align}
The relationship between the incoming field $\phi^{\rm in}(\vr,t)$ and the outgoing one $\phi^{\rm out}(\vr,t)$ can also be worked out within this model exactly and one finds that $B(\vk)=A(\vk)+i\rho(\vk,w(k))$.
As a result the average number of radiated real particles $\bar{N}$ can also be calculated and  indeed $\bar{N}=\bar{n}$, so that they coincide.

This implies that the number of radiated particles matches the initial ($t=0$) budget of virtual particles comprising the cloud around the source mimicking a resonance. Indeed, further insight can be gained by studying the distribution of the virtual particles around the source at $t=0$ by evaluating the expectation value
${_{\rm out}\langle 0}|\phi(\vr,0)|0{\rangle_{\rm out}}$. Its calculation is straightforward by using the relation of Eq.~\eqref{201218.9}, so that
\begin{align}
\label{201218.10}
{_{\rm out}\langle 0}|\phi(\vr,0)|0{\rangle_{\rm out}}&=
\int\frac{d^3k\,g\rho(\vk)e^{i\vk\vr}}{(2\pi)^3 2w(k)}\left\{
\frac{w(k)-M_R}{(w(k)-M_R)^2+\frac{\Gamma^2}{4}}+\frac{w(k)+M_R}{(w(k)+M_R)^2+\frac{\Gamma^2}{4}})
\right\}~.
\end{align}
In the following, to end with algebraic results which is enough for illustrative purposes, we take massless particles and the case of a point source, so that $\rho(\vk)=1$. Then, $w(k)^2=k^2$ and the previous integral can be performed straightforwardly for $r\neq 0$ with the result
\begin{align}
\label{201218.11}
{_{\rm out}\langle 0}|\phi(\vr,0)|0{\rangle_{\rm out}}&=\frac{g}{4\pi r}\cos(M_R r)e^{-\frac{\Gamma r}{2}}~.
\end{align}
This establishes that the budget of virtual particles at $t=0$ are distributed in a region within a distance of order  $2/\Gamma$ around the source at $r=0$. This distance is the one traveled by the particles moving at the speed of light in a time twice the mean lifetime.  Notice that the integration of the modulus squared of the previous wave function is finite, and it  gives
\begin{align}
\label{201218.12}
\int d^3r |{_{\rm out}\langle 0}|\phi(\vr,0)|0{\rangle_{\rm out}}|^2=\frac{g^2}{8\pi\Gamma}\frac{1+\Gamma^2/2M_R^2}{1+\Gamma^2/4M_R^2}~.
\end{align}

\end{document}